\documentstyle[aaspp4,epsfig]{article}
 \def\simlt{\lower.5ex\hbox{$\; \buildrel < \over \sim \;$}}
  \def\simgt{\lower.5ex\hbox{$\; \buildrel > \over \sim \;$}}
\begin{document}
\title{Gravitational Lensing of High Redshift Type Ia Supernovae: \\
A Probe of Medium Scale Structure}

\author{R. Benton Metcalf}
\affil{\it Departments of Physics  and Astronomy, and Center for Particle
Astrophysics \\ University of California, Berkeley, California 94720}

\begin{abstract}
The dispersion in the peak luminosities of high redshift type Ia supernovae 
will change with redshift due to gravitational lensing.  This lensing is 
investigated with an emphasis on the prospects of measuring it and separating it from other 
possible sources of redshift dependent dispersion.  Measuring the lensing induced dispersion 
would directly constrain the power spectrum of density fluctuations on smaller length scales 
than are easily probed in any other way.  The skew of the magnification distribution is related 
to the bispectrum of density fluctuations.  Using cold dark matter models it is found that 
the amount and quality of data needed is attainable in a few years.  A parameterization of 
the signal as a power law of the angular size distance to the supernovae is motivated by 
these models.  This information can be used in detecting lensing, detecting other systematic 
changes in supernovae and calculating the uncertainties in cosmological parameter estimates.
\end{abstract}

\section{Introduction}

There is presently a large effort underway to predict and detect the weak 
gravitational lensing caused by Large Scale Structure (LSS) or the ``cosmic shear'' 
(see \cite{valdes83,mould94,vill96,kais92,kais96}).  Such a measurement would 
constitute a direct probe of the mass density fluctuations on large scales 
irrespective of how light and baryons are distributed.  This would make it possible 
to measure one of cosmology's least well understood processes, how light traces mass 
and whether this is a function of scale.  Gravitational lensing causes 
images to be both magnified and demagnified as well as stretched asymmetrically 
(shear).  Most of the methods proposed for detecting LSS lensing are based on 
measuring the shear in high redshift galaxy images.  Although the distortions in 
the ellipticities of individual galaxies are expected to be small they are distorted 
in coherent ways.  Indications of lensing are sought in the alignment of 
the galaxy images with the assumption that they are not intrinsically aligned.   This 
technique has already been used with great success on galaxy cluster lensing and is presently 
being applied to random fields in an attempt to detect the lensing effects of 
large scale structure.  In this paper a method of detecting lensing directly through 
its magnification rather than shear is proposed.

The study of high redshift supernovae (SNe) is another area of cosmology and astrophysics 
that is seeing a lot of activity.  Type Ia SNe, the brightest type of 
supernova (SN), 
are believed to be caused by the thermonuclear explosion of an accreting white dwarf.
It has been found empirically that the peak magnitude of type Ia's have a 
dispersion of only 0.2 - 0.3 mag in B band.  It has further been found that 
the peak magnitude is related to the width of the SN's light-curve which 
can then be used to reduce the dispersion to about 0.17 mag if one color is used 
\cite{hamuy96} 
and 0.12 mag if multiple colors are used \cite{RPK96}.  In addition to the light-curve 
width, the SN's color and spectral features are related to the peak luminosity 
\cite{BNF97,nugent95}.  It may be possible to reduce this dispersion in the future by 
incorporating additional observables into the correction procedure.  

The uniformity in type Ia SNe combined with their high 
luminosity makes them an excellent tool for doing cosmology.  Using them to measure 
the redshift-luminosity distance relation has recently resulted in tightened 
constraints on the cosmological parameters $\Omega$ and $\Omega_\Lambda$ 
(\cite{Perl97,Perl98,garn98,Riess98}).  There are now systematic searches for Ia SNe at 
high redshift which can reliably discover on the order of ten SNe in a night's 
observing and do spectroscopic followup (\cite{Perl97}).  In addition there are several ongoing 
searches for low redshift SNe.  To date there have been more then 100 type Ia 
SNe discovered with redshifts between $z=0.4$ and $0.97$ and many additional ones 
at lower redshift.

In this paper I concentrate on the lensing produced by dark matter composed of microscopic 
particles.  Matter in macroscopic compact objects can cause microlensing.  However, the known 
populations of stars will not cause a significant number of microlensing events.  The microlensing 
of SNe has been discussed by \cite{SW87} and \cite{LSW88}.  
A SN could also be lensed by one dominant galaxy cluster or individual 
galaxy.  There is the possibility of getting multiple observable images and high 
magnifications (strong lensing) in this case.  It has been suggested that observing 
SNe behind galaxy clusters would be a way of lifting the mass sheet degeneracy 
that exists in shear measurements of the gravitational lensing \cite{kolatt97}.  The 
likelihood 
of a SN at $z=1$ being strongly lensed by a galaxy or cluster is small 
unless they are specifically sought out.  In general a SN will be lensed by 
many galaxies and larger structures, each have a weak contribution to the 
total magnification.  This will increase the dispersion of high redshift SNe 
and decrease the precision of cosmological parameter determinations.  This decrease in 
precision has been investigated by \cite{Frieman97} using analytic methods and by 
\cite{wamb97} using N-body simulations.  \cite{HW98} calculates the lensing of 
point sources using numerical simulations which assume that all matter is in 
unclustered galaxy halos.  \cite{Kant98} (and \cite{Kant95}) does analytic calculations 
of this effect under the assumption that all matter is in compact objects and none of 
them are close enough to the line of sight to a SN to cause significant lensing.  
In this paper the problem will be turned around and we will ask how well the lensing 
itself can be determined from the SN data.

There are several important differences between the lensing of SNe and the lensing of galaxies.
For SNe the signal to noise ratio in each measurement can in fact be greater.  Lensing is 
estimated to contribute about $5$ to $10\%$ to the observed root-mean-squared ellipticity of a 
$z=1$ galaxy.  For SNe the variance in the lensing contribution is comparable 
to the intrinsic variance in the peak magnitude at this redshift.  
In addition the unlensed ellipticity distribution of 
galaxies at high redshift is not known and can not be easily extrapolated from zero 
redshift galaxies.  For this reason lensing must be inferred by correlations between 
galaxies, either between lensed galaxies or between lensed galaxies and foreground 
galaxies.  The result is that the lensing of galaxies measures the 
shear averaged over a finite area on the sky.  This average shear drops rapidly with 
increasing area 
and the signal to noise is reduced to something more like $1\%$ per 
galaxy on the $1\mbox{ deg}^2$ scale.  This is made up for by the large number of 
evaluable galaxies ($\sim 10^5\mbox{ deg}^{-1}$) to the extent that fluctuations in 
the projected mass density at $1-100$~arcmin scales are expected to be detectable in 
the near future.  In contrast, the dispersion in the absolute magnitude of type Ia 
SNe is presumably independent of redshift.  The dispersion can then be measured 
in a low redshift population where lensing is not important.  The lensing of galaxies 
does have the advantage of sources that are generally at higher redshifts where the 
lensing is stronger.  On the other hand, the redshift distribution of faint 
galaxies is not strongly constrained which adds systematic uncertainty.  The redshifts 
of SNe are individually measured.

It is clear that the lensing of SNe and the lensing of galaxies probe different 
scales of density fluctuations for several reasons.  Because the lensing of galaxies 
can only be detected through correlations in their shear, or positions and shear, 
lensing structures that are smaller than roughly the separation between galaxies 
are not detectable.  Since each SN is an independent measure of the 
lensing at a point, not an average over a region on the sky, they will be sensitive to 
smaller scale structures.  Also the lensing of SNe is a direct measurement of 
the magnification which, in the weak lensing limit, is directly related to the mass 
density along the line of sight.  The shear is dependent on the mass outside of 
the ``beam'' and thus a shear map is in a way a smoothed version of a surface 
density map.  In the thin lens approximation the shear and the magnification are related to 
each other through a differential operator \cite{KS93} which makes the shear insensitive to a 
uniform 
offset in the surface density - the ``mass sheet degeneracy''.  
Although the lensing of galaxies has limitations on small scales it does have the 
potential of measuring lensing over a large range of scales, arcminutes to degrees.  
With the possible exception of SNe viewed through galaxy clusters, it will be  
difficult to find enough SNe that are close enough together to measure 
correlations in their lensing.

In the next section I describe first how the lensing of SNe is related to the 
density fluctuations and then I describe how the lensing signal could be identified 
in the data. In section~\ref{models} a specific model of structure formation is used 
to estimate the level of signal and to motivate some parameterizations.  The last 
section contains conclusions and remarks about possible complications.

\section{Formalism}

In section~\ref{LSSlens} I show how the variance and skew of the magnification of SNe 
are related to the power spectrum and bispectrum of density fluctuations in the universe.  
In this section only a few loose and justifiable restrictions on the form of the density 
fluctuations will be required.  In section~\ref{stat} some formal comments are made on 
extracting parameters from the data and on estimating the precision with which parameters 
can be measured.  It is assumed throughout that the lensing is weak and that multiple 
images are not produced.  Less than one in $\sim 10^{-3}$ SNe at $z\sim 1$ are expected to 
be strongly lensed so this assumption is well justified for the redshifts considered here.

\subsection{The Magnification}
\label{LSSlens}

Lensing can be viewed mathematically as a mapping from points on a source 
plane to points on a image plane.  The Jacobian matrix of this mapping can be 
written as the identity plus a perturbation, $J_{ij}(\theta)=\delta_{ij}+
\Phi_{ij}(\theta)$.  In the weak lensing limit the magnification is given by 
\begin{equation}
\mu(\theta)=\det\left[ {\bf I} + {\bf \Phi}(\theta) \right]^{-1} \simeq 1- \mbox{Tr}{\bf \Phi}(\theta)
\end{equation}
It is standard practice to define the convergence, $2 \kappa(\theta) \equiv -\mbox{Tr}{\bf \Phi}(\theta)$.  The change in the magnitude of a point source is then
\begin{equation}
m-\overline{m}=2.5 \log\left[\mu(\theta)\right] \simeq 2.1715 \kappa(\theta).
\end{equation}

To calculate correlations in $\kappa(\theta)$ I will use a 
perturbation method based on the techniques developed in \cite{gunn67,blan91,kais92}.
The shear tensor is given by
\begin{eqnarray}
\Phi_{ij} \equiv  \frac{\partial \delta \theta_i}{\partial \theta_j} = -2 \int^r_0 dr' W(r,r')\phi_{,ij}(r') & ; & W(r,r_s)\equiv \frac{g(r)g(r_s-r)}{g(r_s)} 
\label{deltphi}
\end{eqnarray}
where $r$ is the comoving radial distance and $g(r) = \{ R_c\sinh(r/R_c),r,R_c\sin(r/R_c) \}$ for 
the open, flat and closed global geometries respectively.  The curvature scale is $R_c=|H_o \sqrt{1-\Omega-\Omega_{\Lambda}}|^{-1}$.  Subscripts with commas 
in front of them denote partial derivatives and $\phi(x)$ is the potential.

The convergence will on average be zero, but it will be different for each SN.  We need 
the correlation in the convergence of two sources at points ${\bf \theta}_1$ and 
${\bf \theta}_2$ on the sky and at coordinate distances $r_1$ and $r_2$.
\begin{equation}
\langle \kappa(\theta_1,r_1) \kappa(\theta_2,r_2) \rangle = 
 \int_0^{r_1} dr' W(r_1,r') \int_0^{r_2} dr'' W(r_2,r'') \langle \bigtriangledown_\perp^2\phi(\theta_1,r')\bigtriangledown_\perp^2\phi(\theta_2,r'') \rangle \label{kappa1}
\end{equation}
It is useful to work in Fourier space at this point.  If the potential 
is statistically homogeneous and isotropic the power spectrum of the potential is 
related to its Fourier coefficients by $ P_\phi(k)= (2\pi)^3 \delta^3({\bf k}-{\bf k}') \langle \tilde{\phi}_k \tilde{\phi}_{k'} \rangle$.  The power spectrum of the
potential is then related to the power spectrum of the mass density contrast by 
Poisson's equation, $P_{\phi}(k,\tau)=9 a(\tau)^{-2}\Omega^2_o \mbox{H}_o^4 k^{-4} 
P(k,\tau)/4$ where $a(\tau)$ is the scale factor normalized to 1 at the present epoch.

The expression (\ref{kappa1}) can be significantly simplified by using an 
approximation that is equivalent to Limber's equation \cite{Limber,kais92}.  This is a 
very good approximation for most realistic models of large scale structure. 
With this approximation equation (\ref{kappa1}) can be reduced to
\begin{eqnarray}
\langle \kappa(\theta_1,r_1) \kappa(\theta_2,r_2) \rangle =\left( \frac{3}{2} \Omega_o H_o^2 \right)^2  \int_0^{r_1<r_2} dr' W(r_1,r')W(r_2,r') \label{kappa2} \\ 
\times \int_0^\infty \frac{dk}{2\pi} a(\tau')^{-2} k P(k,\tau') J_o\left( g(r')|\theta_1-\theta_2 | k\right).  \nonumber
\end{eqnarray}
where $J_o(x)$ is the zeroth order Bessel function.  The Bessel function reduces the correlations 
between SNe to insignificant levels unless the SNe are very close together.  The variance is found 
by taking $\theta_1=\theta_2$, $r_1=r_2$.

In a like manner the third moment of $\kappa(r_s)$ can be expressed in terms of the bispectrum. 
In the case of a homogeneous and isotropic field the bispectrum can be defined by 
\begin{eqnarray}
\langle \tilde\delta({\bf k})\tilde\delta({\bf k'})\tilde\delta({\bf k''}) \rangle = (2\pi)^3\delta^3( {\bf k} + {\bf k'}+{\bf k''})\, B(k,k',\theta)
\end{eqnarray}
where $\theta$ is the angle between the vectors ${\bf k}$ and ${\bf k'}$.  Correlations between 
sources will be even smaller in this case so I consider only the one point correlation.
With similar assumptions to those used in getting (\ref{kappa2}), the third moment is
\begin{eqnarray}
\langle \kappa(r_s)^3 \rangle = 2 \left[ \frac{3 \Omega_o H_o^2}{4\pi} \right]^3 \int^{r_s}_0 
\left( \frac{W(r,r_s)}{a(r)}\right)^3 \int_0^\infty d k \int_0^\infty d k' \int_0^\pi d \theta ~ 
k k' B(k,k',\theta,\tau). \label{kappa3}
\end{eqnarray}
The bispectrum on large scales (small $k$) is expected to be small or zero.  The third and higher 
moments of $\kappa$ are dependent on significantly smaller scale structure than the second moment.
Note that no assumption of Gaussianity or linear evolution of structure has been used in deriving (\ref{kappa2}) or (\ref{kappa3}).

\subsection{Measuring Parameters}
\label{stat}

The question addressed here is how to measure the parameters of a model 
using the SN data.  Conversely, we would also like to know the amount and quality of 
the data that will be necessary to measure parameters to a given accuracy.  To do 
these things I use a likelihood analysis approach.  As seen in the last section the distribution of $\kappa$ is not expected to 
be Gaussian.  Nonlinear clustering causes the distribution to be skewed in favor of 
demagnification.  This is simply because the mass is concentrated into small regions so 
a typical line of sight will tend to travel disproportionately through underdense 
regions.  The distribution of corrected, intrinsic magnitudes may not be Gaussian either. 
At present the low redshift SNe show some evidence of skewness, but are roughly 
consistent with Gaussian after corrections for extinction.  Extinction corrections are more difficult for high redshift SNe so non-Gaussianity may be introduced in this way as well.
However, with the observational errors included one might expect the distribution of 
$\Delta m$ to be close to Gaussian because it is the result of several independent random processes.
For the remainder of this paper I will make the simplifying assumption that $\Delta m$ is 
Gaussian distributed.

The logarithm of the Gaussian likelihood function is
\begin{eqnarray}
{\ln\cal L}& =& -\frac{1}{2} \left[ {\bf\Delta m^T} {\bf C}^{-1} {\bf\Delta m} + \mbox{Tr}\ln{\bf C} \right]
\label{likelihood}\\
&=& -\frac{1}{2} \sum_i \left[ (\Delta m_i)^2 \left(4.715 \langle\kappa_i^2\rangle + (\sigma_m^2)_i\right)^{-1} + \ln\left(4.715 \langle\kappa_i^2\rangle + (\sigma_m^2)_i \right) \right] \label{diagonal}
\end{eqnarray}
where $\Delta m_{i} \equiv m_i - \overline{m}_i$, the difference between the observed 
and expected magnitudes, and $\bf{C}$ is the covariance matrix as 
predicted by the model.  Expression~(\ref{diagonal}) is for the special case where there are no correlations between SNe.  The subscripts correspond to each of the SNe observed. 
I will divide the covariance matrix into two parts, the first due to 
lensing and the second due to other sources of  dispersion in SN peak 
luminosities - $C_{ij} =4.715 \langle\kappa_i \kappa_j\rangle(z) + (\sigma_m^2)_i \delta_{ij} + C^{cor}_{ij}$.  The matrix $C^{cor}_{ij}$ has no nonzero diagonal elements and is 
included to account for correlations between the SN luminosities that might arise 
from the light-curve and/or color corrections or correlations that might arise between SNe 
observed on the same night or with the same telescope for example.  The diagonal elements of 
$C_{ij}$ are the $\sigma_i$'s.  In turn we can divide $(\sigma_m^2)_i = \sigma_M^2 + 
(\sigma_n^2)_i$ where $\sigma_M$ is the variance in the intrinsic peak magnitude which is the same 
for all SNe and $(\sigma_n^2)_i$ is the variance due to noise, light-curve fitting, etc.  There 
is no sum over repeated indices.  Maximizing (\ref{likelihood}) or (\ref{diagonal}) with respect 
to all the model parameters will result in the set of parameters that best fits the data.  
Constraints from other, independent determinations of parameters could be incorporated into the 
likelihood function with a prior distribution.

Now consider the problem of estimating the precision with which cosmological 
or model parameters will be determined.  The precision can be estimated by the 
ensemble average of the Fisher matrix which is made of the second derivatives of the 
likelihood function with respect to those parameters.  If we consider two parameters $\alpha$ 
and $\beta$,
\begin{eqnarray}
\langle -\frac{\partial^2 \ln {\cal L}}{\partial\alpha\partial\beta} \rangle &=&
\langle (\alpha-\overline{\alpha})(\beta-\overline{\beta}) \rangle_{\cal L}^{-1} =
\mbox{Tr}\left[ \overline{{\bf m}},_\alpha  \overline{{\bf m}},_\beta^T
 {\bf C}^{-1} + \frac{1}{2} 
{\bf C}^{-1}{\bf C},_\alpha {\bf C}^{-1} {\bf C},_\beta \right] \label{precision} \\
&=&\sum_i \left( 4.715 \langle\kappa_i^2\rangle + (\sigma_m^2)_i \right)^{-2} \left\{ \overline{m_i},_\alpha \overline{m_i},_\beta (4.715 \langle\kappa_i^2\rangle + (\sigma_m^2)_i ) + \frac{(4.715)^2}{2} \langle\kappa_i^2\rangle,_\alpha \langle\kappa_i^2\rangle,_\beta \right\} \nonumber
\end{eqnarray}
The last line is true when the SNe are statistically independent.
The first terms in these expressions are the usual term that comes from uncertainties 
in measuring the mean magnitude of the SNe and the second term is the 
result of the dispersion of the magnitudes, due to lensing or other causes, being a 
function of the parameters being measured. 

\section{Scalings, Models and Predictions}
\label{models}

To make some quantitative and qualitative estimates I use models whose mass density 
is dominated by cold dark matter (CDM).  It is helpful to first consider what 
might be expected for the form of the lensing signal.  In linear theory $P_{L}\left(k,a(\tau)\right) \propto \left[a(\tau)g\left(\Omega(\tau),\Omega_\Lambda(\tau)\right)\right]^2$ where $g(\Omega,\Omega_\Lambda)$ is given in \cite{CPT}.  This allows the integral in (\ref{kappa2}) to decouple so that factors dependent on the power spectrum can be separated from 
factors dependent on cosmological parameters.  The same thing happens in the opposite 
extreme when structure is highly nonlinear and fully virialized.  In this case structure is 
stable in real space and the power spectrum evolves like $P_{NL}\left(k,a(\tau)\right) \propto a(\tau)^3$ regardless of cosmological parameters.  These two extreme cases can be explicitly calculated for the $\Omega=1$ model where $P_{L}\left(k,a(\tau)\right) \propto a(\tau)^2$,
\begin{equation}
\langle \kappa(z)^2 \rangle = \frac{3}{40} \left(r(z)H_o\right)^3 H_o \int_0^{\infty}\frac{dk}{2\pi}kP(k) \times
 \left\{ \begin{array}{cl} 
1 &  \mbox{linear evolution}\\ 
1-\frac{1}{2}r(z)H_o + \frac{1}{14}\left[r(z)H_o\right]^2 & \mbox{stable evolution}
\end{array} \right. \label{estimate}
\end{equation}
where the power spectrum is evaluated at the present day.  These two results bracket the 
real result for $\Omega=1$ models - linear evolution bounds it an the top and stable 
evolution on the bottom.  The same thing can be done for $\Omega \neq 1$ models.  In this 
way if $\langle \kappa^2 \rangle$ is measured firm bounds on the integral of $kP(k)$ can 
be found within a model.  The model will already be strongly constrained by 
$\overline{m}(z)$ and other observations.  If the redshift dependence of $\langle \kappa^2 \rangle$ could be firmly established insight into the evolution of clustering would be 
gained.  For $\Omega \neq 1$ models $P_{L}\left(k,a(\tau)\right)$ is a somewhat steeper 
function of $r(z)$ because of the decay of potential fluctuations.  The stable clustering 
case is less strongly dependent on $\Omega$ except for the $\Omega^2$ factor in (\ref{kappa2}) which comes from Poisson's equation.  The geometric factors in $\Omega + 
\Omega_\Lambda \neq 1$ models will tend to make $\langle\kappa(r)^2\rangle$ steeper, all 
other things being equal.

\begin{figure}
\centering\epsfig{figure=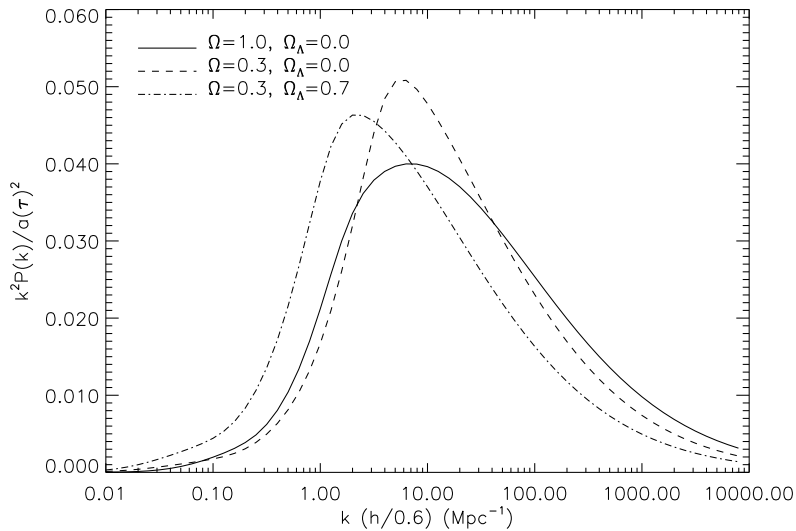,height=3in}
\caption[kpk.eps]{The scale dependence of the lensing of point sources in CDM models.}
\label{power}
\end{figure}

To make things more quantitative I use the linear power spectrum of matter 
fluctuations given by \cite{sugy95}:
\begin{eqnarray}
&P(k)=Ak^n T(k e^{\Omega_b+\Omega_b/\Omega}/\Omega h^2)^2 &
\\
&T(q)=\frac{\ln(1+2.34q)}{2.34q}\left[ 1+3.89q+(16.1q)^2+(5.46q)^3+(6.71q)^4 \right]^{-1/4}& \nonumber
\end{eqnarray}
To convert this to a nonlinear power spectrum I use the technique of \cite{peac96} 
which is based on N-body simulations and thus does not take into account any 
hydrodynamics that could be important on small scales.  In all calculations 
$\Omega_b = 0.015 h^{-2}$.  This power spectrum goes to 
$k^{-3}$ at small scales so that the dimensionless power spectrum $k^3P(k)$ becomes 
scale independent.  Figure~\ref{power} shows what range in $k$-space is responsible 
for lensing in these models.  The power spectrum in this range is speculative because 
the \cite{peac96} formulae which are fits to N-body simulations become less certain at 
small scales.
I choose to cut the power spectrum off at $k=1000\mbox{ Mpc}^{-1}$ for the purposes 
of (\ref{kappa2}).  A cutoff at $k=100\mbox{ Mpc}^{-1}$ reduces $\langle\kappa^2\rangle$ 
by $\sim 10 - 18\% $.

\begin{figure}[t]
\centering\epsfig{figure=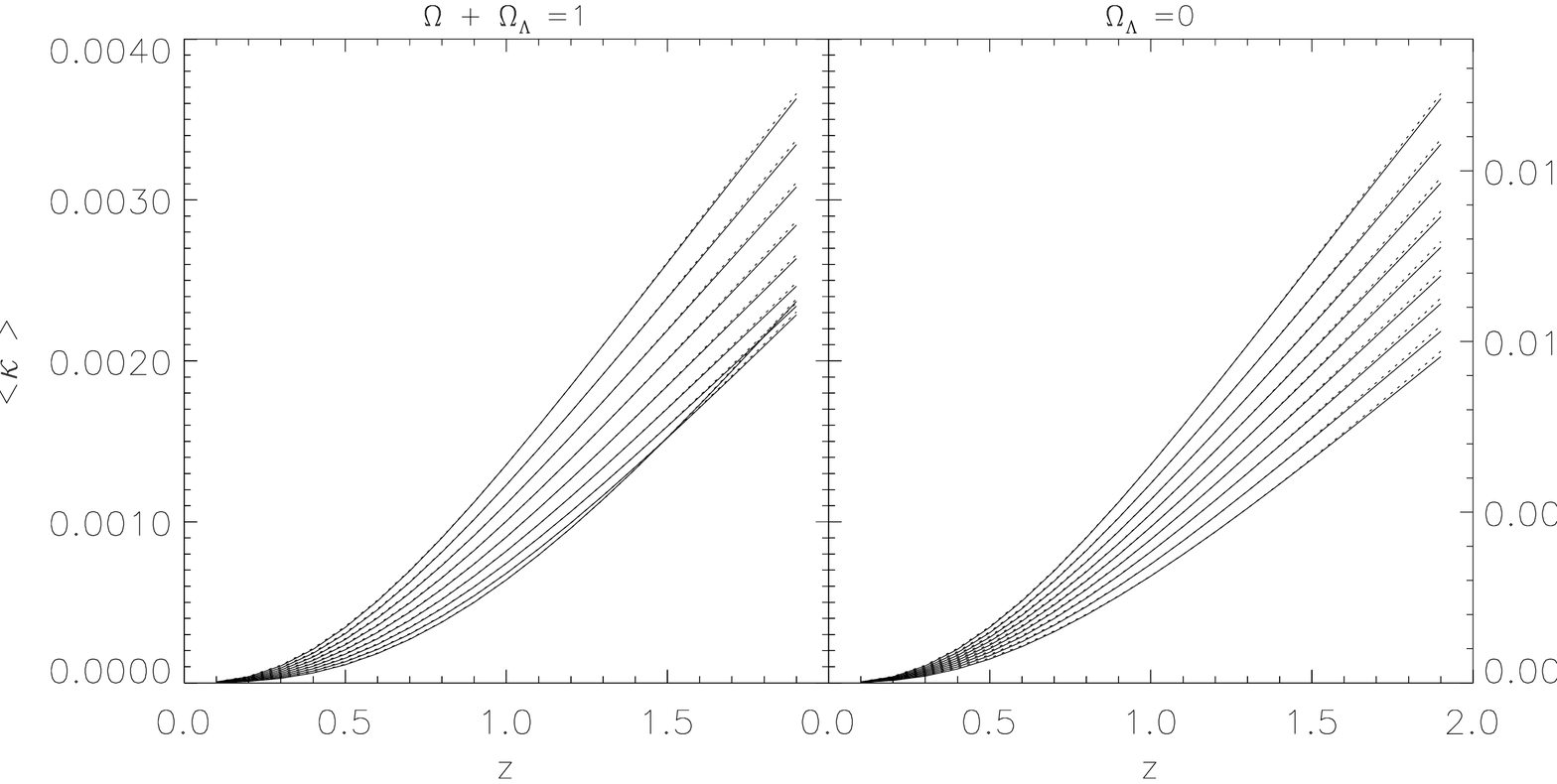,height=3.5in}
\caption[kappaSQR.eps]{The second moment of $\kappa$ in CDM models.  On the left are flat 
models and on the right, open models.  The dotted curves are the fits given in the text. 
The top curve in each plot is for $\Omega=1$.  Each successive curves going down has 
$\Omega$ reduced by $0.1$ from the one above it.  All the models have the normalization that best fits galaxy cluster abundances and $h=0.6$.}
\label{kappas}
\end{figure}
Motivated by (\ref{estimate}) I find a convenient fit to the variance of the convergence in these models
\begin{equation}
\langle \kappa(z)^2 \rangle \simeq \eta_o^2 \left[ r(z,\Omega,\Omega_\Lambda)H_o \right]^{\gamma}
\end{equation}
where $r(z,\Omega,\Omega_\Lambda)$ is the comoving distance.
This fits all the $k<1000\mbox{ Mpc}^{-1}$, cluster-normalized CDM models reasonably well 
with $\gamma \simeq 2.92$ for flat models and $\gamma \simeq 2.80 + 0.11\Omega $ for 
$\Omega_\Lambda=0$ models as is shown in figure~\ref{kappas}.   The constant $\eta_o$ depends on 
how the power spectrum is normalized.  If it is normalized to cluster 
abundances using the formulae of \cite{VL96} with $h=0.6$ I get 
$\eta_o^2 \simeq (0.70+1.44 \Omega^2)h\times 10^{-2}$ for $\Omega_\Lambda=0$ models 
and $\eta_o^2 \simeq (0.32+1.80 \Omega^2)h\times 10^{-2}$ for flat models.  In general the convergence does not have a simple 
dependence on $H_o$ because both the power spectrum's normalization and shape are 
dependent on it.  If the normalization is kept constant and $\Omega$ is varied 
independently, $\eta_o$ is a steeper function of $\Omega$, but this normalization 
will always be 
inconsistent with other observations for some range of $\Omega$.  To properly 
incorporate independent constraints on the normalization (or any other parameter), a 
prior distribution should be incorporated in the likelihood function, 
(\ref{likelihood}).  As is expected from the discussion above, $\langle \kappa(z)^2 
\rangle$ falls somewhere between stable and linear evolution for all the models.
Here the lensing parameter $\eta_o$ is estimated using CDM models, but this 
is only for the purposes of prediction.  With data $\eta_o$ can be measured.  The value of 
$\eta_o$ is essentially a measure of the power spectrum on small scales.  The 
power spectrum is not very well constrained on these scales by any method that can really be 
called unbiased.

\begin{figure}[t]
\centering\epsfig{figure=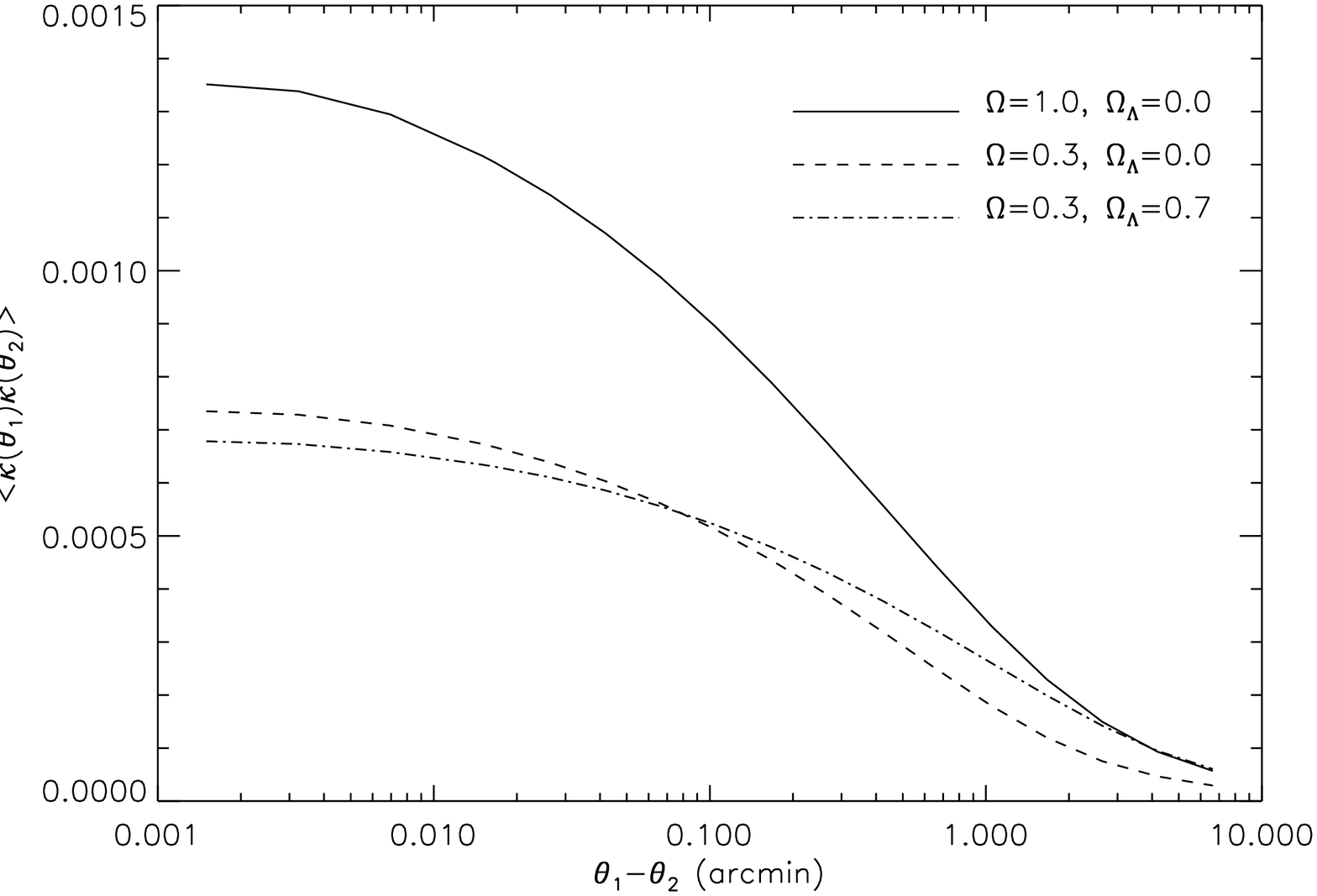,height=2.5in}
\caption[kappa_s.eps]{The second moment of the convergence as a function of angular 
separation between sources.  All cases are for cluster normalized CDM models with $h=0.6$ 
and the sources are at $z=1$.}
\label{kappa_thetas}
\end{figure}
Figure~\ref{kappa_thetas} shows the angular dependence of $\langle \kappa(\theta_1)\kappa(\theta_2) \rangle$.  The correlations between SNe will be very small if 
they are separated by more than about an arc-minute.  It will be difficult to find enough 
SNe that are close enough together for this cross-correlation to be 
measured.  The large angular sizes and high densities of galaxy clusters make them an 
exception to this role, but a random line of sight is not likely to pass through a 
cluster.  If these cross-correlations could be measured it would be sensitive to 
density structures of a larger scale then the diagonal elements, $\langle \kappa^2_i \rangle$.

\begin{deluxetable}{cccccrrrcrrrcrrr}
\tablecolumns{16}
\tablewidth{0pt}
\small
\tablecaption{Number of SNe Needed for Detection \label{table}}
\tablehead{
\multicolumn{4}{c}{Model}&\colhead{} & \multicolumn{3}{c}{$z=0.5$} &\colhead{}& \multicolumn{3}{c}{$z=1.0$} &\colhead{}& \multicolumn{3}{c}{$z=1.5$}\\
\cline{6-8}  \cline{10-12}\cline{14-16} \\
$\Omega$ & $\Omega_\Lambda$ & $\sigma_8$& $h$ &\colhead{} & $\Delta m$ & $N_{0.10}$& $N_{0.13}$ &\colhead{}& $\Delta m$& $N_{0.10}$& $N_{0.13}$&\colhead{} & $\Delta m$& $N_{0.10}$& $N_{0.13}$}
\startdata
1.0 & 0.0 & 0.6 & 0.60 && 0.04 & 103 & 261 && 0.08 & 13 & 27 &&  0.11 & 7 & 11 \nl
1.0 & 0.0 & 0.6 & 0.75 && 0.05 & 62 & 153  && 0.09 & 9 & 17  && 0.13 & 5 & 8 \nl
1.0 & 0.0 & 0.5 & 0.65 && 0.03 & 252 & 668 && 0.06 & 26 & 59 && 0.09 & 11 & 22 \nl
1.0 & 0.0 & 0.8 & 0.65 && 0.06 & 33 & 76 && 0.11 & 6 & 10 &&  0.16 & 4 & 5 \nl

0.3 & 0.0 & 1.0 & 0.60 && 0.03 & 370 & 995 && 0.06 & 30 & 69 &&  0.08 & 11 & 23 \nl
0.3 & 0.0 & 1.0 & 0.75 && 0.03 & 261 & 694 && 0.06 & 23 & 50 &&  0.09 & 9 & 17 \nl
0.3 & 0.0 & 0.7 & 0.65 && 0.02 & 1456 & 4031 && 0.04 & 97 & 246 &&  0.06 & 31 & 70 \nl
0.3 & 0.0 & 1.4 & 0.65 && 0.04 & 89 & 223 && 0.09 & 10 & 20 &&  0.13 & 5 & 8 \nl

0.3 & 0.7 & 1.2 & 0.60 && 0.02 & 592 & 1611 && 0.06 & 34 & 79 &&  0.08 & 11& 22 \nl
0.3 & 0.7 & 1.2 & 0.75 && 0.03 & 432 &1165 && 0.06 & 26 & 59 &&  0.09 & 9 & 18 \nl
0.3 & 0.7 & 0.8 & 0.65 && 0.02 & 2419 & 6748 && 0.04 & 114 & 291 &&  0.06 & 31 & 72 \nl
0.3 & 0.7 & 1.6 & 0.65 && 0.04 & 139 & 361 && 0.08 & 11 & 23 &&  0.13 & 5 & 8\nl

\tablecomments{These are $2\sigma$ detection limits.  $N_{0.10}$ and $N_{0.13}$ refer to 
the number of SNe needed if $\sigma_m = 0.10$ and $0.13$.}
\enddata
\end{deluxetable}

To estimate how well $\eta_o$ can be measured we can use (\ref{precision}) to find
\begin{equation}
\langle (\eta_o - \langle \eta_o \rangle)^2 \rangle^{-1}_{\cal L} = 
\frac{1}{2} \sum_i \left[ \frac{9.43 \eta_o (r_i H_o)^\gamma}{4.715 \eta_o( r_i H_o)^\gamma + \sigma_m^2} \right]^2.
\end{equation}
Table~\ref{table} gives estimates of the numbers of SNe needed to make a 
$2\sigma$ detection of $\langle\kappa^2\rangle$ calculated using (\ref{precision}).  
The SNe are taken to all be at the same redshift, $z=0.5$, $z=1.0$ and 
$z=1.5$.  The numbers $N_{0.10}$ and $N_{0.13}$ are for $\sigma_m = 0.10$ and $0.13$.  
The number goes as $\sigma_m^4$ so it is very sensitive to this parameter.  The models 
are intended to span the range that is consistent with cluster abundances and 
determinations of $H_o$.

\begin{figure}[t]
\centering\epsfig{figure=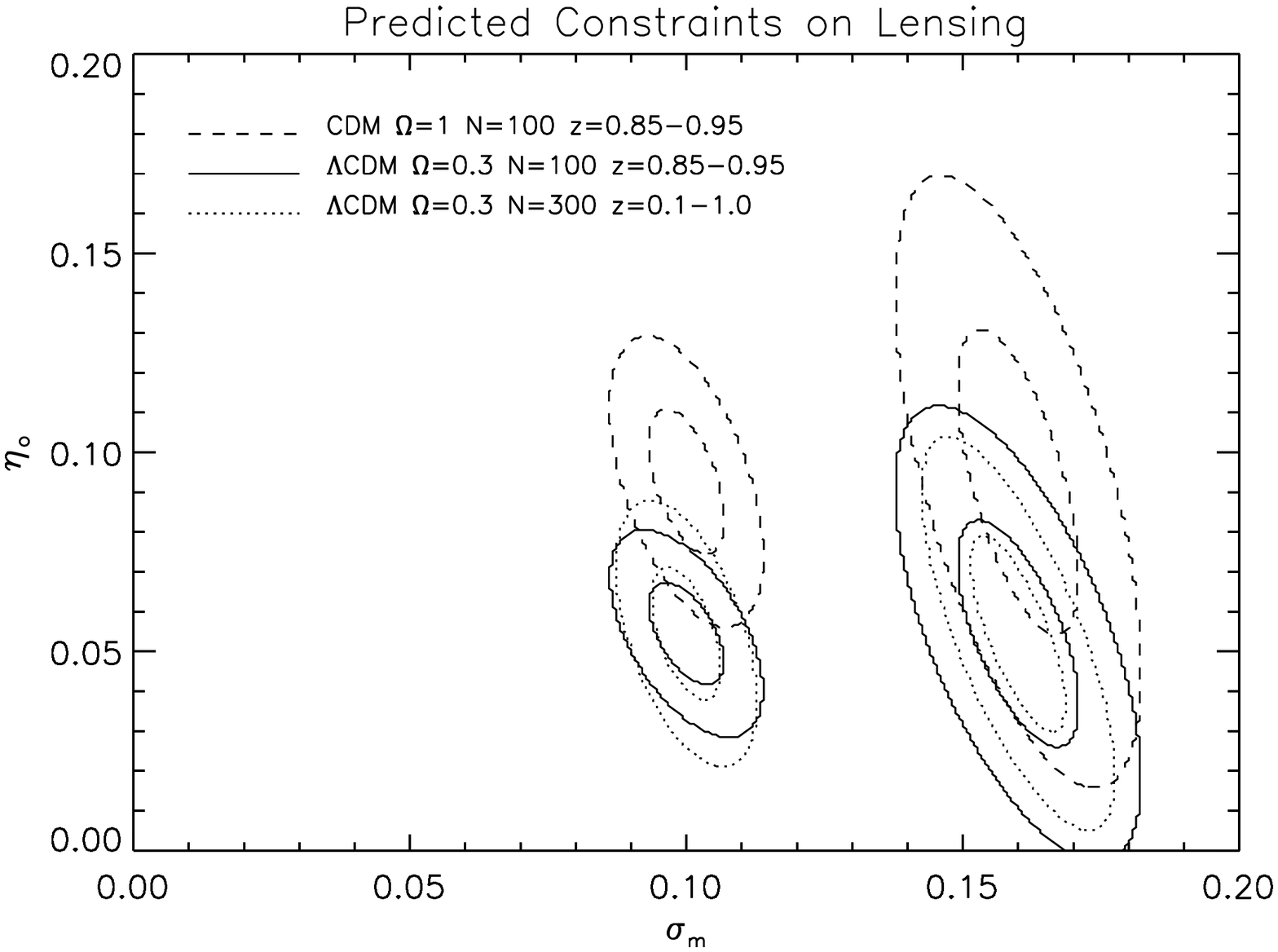,height=3.5in}
\caption[lnerrorV.eps]{The estimated confidence regions for the combination of lensing and 
``intrinsic'' variance in the SNe magnitudes.  The number of high redshift SNe, $N$, are listed 
in the upper left along with the redshift range where they are observed.  They are assumed 
to be distributed in redshift according to comoving volume within this range.  In addition to the 
high redshift SNe there are assumed to be 100 ``low'' redshift SNe in each case which are not 
lensed.  The inner ellipses are the estimated $68\%$ confidence regions and the outer the $95\%$.  
The two cases for each model are with assumed $\sigma_m=0.1$ and $0.16$ on the left 
and right respectively.
} 
\label{likeliplot}
\end{figure}

Of course the SNe will not all be at the same redshift and the ``intrinsic'' 
variance, $\sigma_m$, in the SN magnitudes will not be perfectly determined.  
It makes sense to try to determine both $\sigma_m$ and $\eta_o$ at the same time.  
The Fisher matrix (\ref{precision}) can be used to calculate an estimated error matrix.  
The result can be represented by an ellipsoid is parameter space. 
This is plotted in figure~\ref{likeliplot}.  The redshift distribution of the SNe is taken to be proportional to comoving volume within the redshift ranges listed.
This is an idealization which assumes that the observed SN rate and the detection 
efficiency are not functions of redshift.  A separate population of 100 $z \simeq 0$ 
SNe is also included.  These are found by searches that target local galaxies as opposed to 
searches for high redshift SNe which can be done either by cataloging galaxies or by 
differencing multiple observations of more or less random fields.
Changing the $\sigma_8$ normalization within observational constraints has a significant 
effect on the number of SNe required to measure $\eta_o$.
The $\sigma_8$'s listed are for the linear power spectra.  Changes in 
the linear normalization are magnified in the nonlinear power spectrum 
so that they show up strongly in the lensing.  
Low $\Omega$ models 
have significantly lower $\eta_o$'s than high $\Omega$ with the same linear normalization 
or with cluster normalization which increases with lower $\Omega$.  This is largely 
because of the factor of $\Omega^2$ in $\langle \kappa^2 \rangle$.  Getting 
to high redshift will be important for discriminating between models because the higher 
the redshift the less elongated the likelihood contours are along the $\eta_o$ axis.
Fifty SNe at $1<z<1.5$ do substantially better at constraining $\eta_o$ than do 100 a $z=0.9$.
Figure~\ref{likeliplot} also shows that simple estimates, like the ones in 
table~\ref{table}, that do not take into account the fact that our knowledge of $\sigma_m$ is 
limited can give deceptive answers.  More sources are needed to detect lensing when $\sigma_m$ 
is not well known because increasing $\sigma_m$ can make a lower $\langle \eta_o^2 \rangle$ 
acceptable.

The rate at which high redshift SNe can be discovered is fundamentally limited by the rate 
at which they occur, the area on the sky that is surveyed and the efficiency with 
which they can be detected.  Although the rate of star formation at high redshift 
can be estimated from observations \cite{madau96}, the type Ia SNe are expected 
to lag significantly behind it.  The amount of time required for a white dwarf to 
accrete enough material to go SN is not only unknown, but probably varies 
greatly on a case to case basis.  There is one measurement of the Ia SN rate at $z 
\sim 0.4$ which is $34.4^{+23.9}_{-16.2} \mbox{ yr}^{-1}\mbox{deg}^{-2}$ 
\cite{pain96}.  If the rest frame rate per comoving volume remains constant the 
observed rate of SNe at $z$ is $R(z)=R_o (1+z_o)g\left(r(z)\right)^2/(1+z) 
g\left(r(z_o)\right)^2$ which makes it 
2 or 3 times larger at $z=1$.  More thoughtful estimations predict that the rest frame 
rate per comoving volume will go up by a factor of 1 to 2.5 \cite{sad97,RC98}.  
A SN must be detected within about a one week window in order for it to be 
usable.  So it is reasonable to estimate that there are $\sim 20-50$ usable type Ia 
SNe per $\mbox{deg}^2$ below a redshift $1.5$ at any given time.
This makes detecting hundreds of high redshift SNe possible within a few years 
if difficulties with spectroscopic confirmation and k-corrections can be overcome.\footnote{The Supernova Cosmology Project (\cite{Perl97}) currently surveys $\sim 3\mbox{ deg}^2$.}

These calculations of $\langle \Delta m^2 \rangle$ agree well with those of
\cite{Frieman97}.  The numerical simulations of (\cite{wamb97,wamb98}) give somewhat 
smaller values.  This is probably due to the combination of their using the COBE 
normalization which is smaller then the cluster normalization for low $\Omega$ models 
and their simulation having a resolution of $\sim 13 h^{-1}$~kpc ($k \sim 480 h\mbox{ Mpc}^{-1}$).

\section{Conclusions}

It has been shown that measuring the gravitational lensing of type Ia SNe is 
feasible if the noise can be reasonably constrained.  It would be best to solve for 
the best fit cosmological parameters (ie. $\Omega$, $\Omega_\Lambda$), lensing 
strength (ie. $\eta_o$) and intrinsic noise ($\sigma_M$) simultaneously using 
SNe at all redshifts.  The photometric uncertainties should be comparatively 
well determined for each SN.  One could then marginalize over 
the intrinsic variance, $\sigma_M$.

The greatest worry is of course that the type Ia SNe 
properties or their galactic environments are changing with redshift.  Observations of 
spectral features and colors (\cite{Perl97}) suggest that this is not the case, but the possibility 
remains.  It is possible that a systematic change in the metallicity of progenitors could 
change both the average peak luminosity and its dispersion.  Another worry is that
extinction corrections change with redshift.  This could systematically reduce $\overline{m}(z)$, 
increase its dispersion and make its distribution non-Gaussian.  Extinction should be accompanied 
by reddening, which can be detected, but the extinction law is not certain.
These changes would affect cosmological parameter parameter estimates as well as lensing 
estimates.  The methods discussed here can be directly applied to detecting any redshift 
dependent change in the dispersion.  In testing for possible redshift evolution 
lensing must be incorporated.

There are difficulties in searching for SNe at higher redshifts.  The region of the 
spectrum that is used to do the light curve correction passes out of the visible at 
$z\gtrsim 1$.  To go to significantly higher redshift may require switching to the IR.  
There is also difficulties with the K-correction and the subtraction of atmospheric lines.  
But with this in mind it seems that gravitational lensing of SNe could be detectable in the 
next few years when hundreds of high redshift SNe are observed and systematic effects are better 
understood.  CCD cameras with fields of view approaching a square degree and very small pixel sizes 
are being built now.  They will be used for weak lensing measurements using galaxy shear. 
 SNe searches could be incorporated into these surveys with the benefits of 
improved cosmological parameter estimation and complimentary weak lensing measurements.

Combined with the limits on the cosmological parameters the lensing of SNe can 
constrain the power spectrum of the true mass density, unbiased by the light 
distribution, on the scale of galaxy halos.  At present the mass distribution on these 
scales is not well known with the exception of within galaxy clusters which are certainly 
atypical regions.  In this paper it has been assumed that the the large majority of 
matter in the universe is in the form of WIMPS or some other small particle.  If the 
dark matter is in compact objects like MACHOs the distribution of magnifications will 
be different.  In this way the lensing of SNe 
also provides information on the composition of dark matter.
In addition to the variance of the magnification distribution the 
skewness would provide an important constraint on the nature of structure in the 
nonlinear regime.  A future paper will treat this subject in 
more detail and relate the magnification distribution to the nature of dark matter and 
the structure of galaxy halos.

\acknowledgments

I would like to thank J. Silk, A. Jaffe, S. Perlmutter and R. Pain for very useful 
discussions.  This work was financially supported by NASA.

\end{document}